\documentclass[aps,prl,twocolumn,floatfix]{revtex4}
\usepackage{graphicx}
\usepackage{amssymb}
\usepackage{xspace}
\usepackage{ulem}
\usepackage[breaklinks=true,colorlinks=true,linkcolor=blue]{hyperref}

\begin{document}
\newcommand{\qweak}{$Q_{\rm weak}$\xspace}
\newcommand{\fp}{Fabry-P\'{e}rot\xspace}

\title{A test of local Lorentz invariance with Compton scattering asymmetry}

\author{Prajwal Mohanmurthy$^{1,2,}$\footnote{Currently at Paul Scherrer Institute, 5232 Villigen PSI, Switzerland}}
\author{Amrendra Narayan$^{1,}$\footnote{Currently at Indian Institute of Technology Bombay, Mumbai, India}}
\author{Dipangkar Dutta$^1$}
\affiliation{$^1$Mississippi State University, Mississippi State, MS 39762, USA}
\affiliation{$^2$Massachusetts Institute of Technology, Cambridge, MA 02139, USA}


\begin{abstract}
  We report on a measurement of the constancy and anisotropy of the speed of light relative to the electrons in photon-electron scattering. We used the Compton scattering asymmetry measured by the new Compton polarimeter in  Hall~C at Jefferson Lab
  to test for deviations from unity of the vacuum refractive index ($n$). For photon energies in the range of 9 - 46 MeV, we
  obtain a new limit of $1-n < 1.4 \times 10^{-8}$. In addition, the absence of sidereal variation over the six month
  period of the measurement constrains any anisotropies in the speed of light. These constitute the first study of
  Lorentz invariance using Compton asymmetry. Within the minimal standard model extension framework, our result yield limits
  on the photon and electron coefficients $\tilde{\kappa}_{0^+}^{YZ}, c_{TX}, \tilde{\kappa}_{0^+}^{ZX}$, and $c_{TY}$.
  Although, these limits are several orders of magnitude larger than the current best limits, they demonstrate the
  feasibility of using Compton asymmetry for tests of Lorentz invariance. Future parity violating electron scattering
  experiments at Jefferson Lab will use higher energy electrons enabling better constraints. 

\end{abstract}

\maketitle

\section{Introduction}
Lorentz invariance (LI) is a corner-stone of the two accepted field theoretical descriptions of 
nature; The standard Model (SM) of strong and electro-weak interactions and general relativity (GR). 
Furthermore, Lorentz symmetry is closely linked with charge conjugation, parity, and time reversal symmetry 
(CPT symmetry) via the CPT theorem: any Lorentz invariant field theory must be CPT invariant~\cite{cpt-theorem1,cpt-theorem2}. 
It has also been demonstrated that in any field theory if CPT symmetry is violated Lorentz symmetry must 
also be violated~\cite{greenberg02}. Thus a test of LI is a test of CPT invariance and 
any violation would imply new physics. It is commonly regarded that both SM and GR are low-energy limits of
a more fundamental theory that consistently merges the two at the Plank-scale. 
It is well known that many of the contenders for this more fundamental theory allow for 
small spontaneous violations of Lorentz symmetry~\cite{strings1,strings2,strings3,strings4,strings5}. The experimental search for Lorentz symmetry
violations is thus a search for evidence of Planck-scale physics~\cite{cptandlorentz}. 
  
The close link between CPT symmetry and Lorentz symmetry provides additional motivations to search for 
their violations. This is because the breakdown of CPT symmetry along with baryon number violation, 
could be the source of the baryon asymmetry of the universe. Unlike scenarios involving only CP 
violation, baryogenesis based on CPT violation does not require the breakdown of thermal 
equilibrium~\cite{cptasym1,cptasym2,cptasym3}. Moreover, observations that the expansion of the universe is accelerating~\cite{accexp1,accexp2} have led to the
idea that the universe is dominated by dark energy, suggesting that a new field permeates all space~\cite{darkeng1,darkeng2,darkeng3,darkeng4,darkeng5}. An
interaction of this field with matter  would manifest itself as an apparent breaking of Lorentz symmetry~\cite{cosm}, and
searches for Lorentz symmetry violation can constrain these cosmological models.    

At currently attainable energies, small violations of Lorentz symmetry can be described by an effective field 
theory known as the standard model extension (SME)~\cite{sme951,sme952,sme953}.  The SME is the usual Lorentz invariant 
standard model plus all possible Lorentz symmetry violating terms that can be formed with the 
known fields of the SM and gravity coupled to constant or slowly varying background 
fields, such that the Lagrangian remains Lorentz invariant~\cite{sme951,sme952,sme953}. The SME provides 
a basis for analyzing experiments that search for Lorentz symmetry violation in a wide range of physical 
systems~\cite{litests}. In the SME framework, spontaneous Lorentz symmetry breaking would result in a background
fields which permeates the universe. This background field will change its direction over the course of a day
due to the Earth's rotation, enabling experiments to detect the Lorentz symmetry violation~\cite{kotSA}. 
\begin{figure*}[t!]
\includegraphics*[height=17cm,angle=-90]{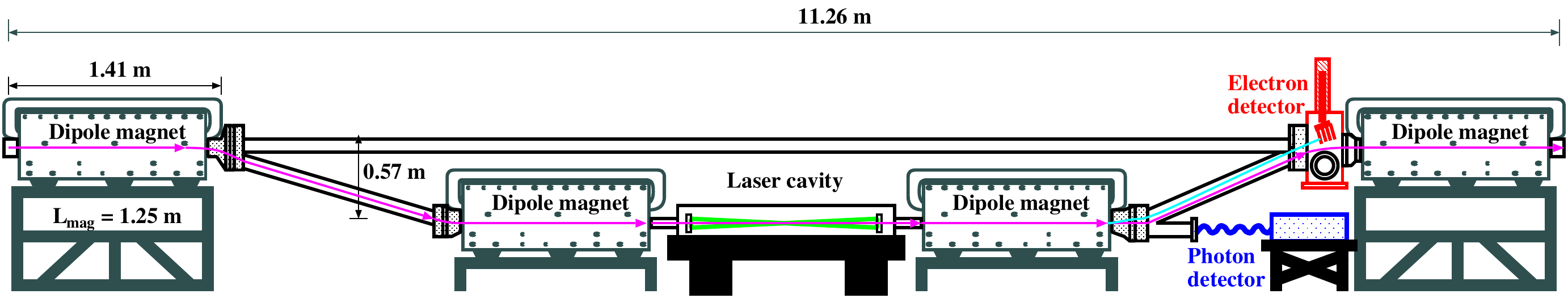}
\caption[]{Schematic of the JLab Hall~C Compton polarimeter.} 
\label{fig:fig1}
\end{figure*}

LI implies that the velocity of light in vacuum, $c(\omega)$, is independent of the energy $\omega$ of the photon. Therefore, tests of isotropy and constancy of the speed of light in vacuum are tests of LI. Such tests were critical in the conception and early experimental verification of LI~\cite{olitests}. Any hypothetical energy dependence of $c(\omega)$ is constrained by the current limit on the photon mass ($m_{\gamma} < $ 10$^{-18}$ eV~\cite{pdg}) as $1 - c(\omega)/c \leq$ 10$^{-36}\omega^{-2}$ eV$^2$, where $c$ is the velocity of a mass-less photon in vacuum. However, the aforementioned Lorentz symmetry breaking background field can influence the propagation of light~\cite{cosm}. In order to distinguish between effects of photon mass and vacuum properties it is more convenient to use vacuum refractive index, $n(\omega) = c/c(\omega)$. 

In an environment where gravity can be neglected, Lorentz violation can be 
described  by the QED limit to the flat-spacetime minimal standard model extension (MSME)~\cite{sme951,sme952,sme953,hohensee091,hohensee092,hohensee093}. In the MSME framework the photon's dispersion relation is modified as:
\begin{equation}
\omega =  (1 - \vec{\kappa}\cdot{\hat k})k + \mathcal{O}(\kappa^2),
\label{eq:disp}
\end{equation}
where $k^{\mu} = (\omega, k{\hat k})$ denotes the photon's four-momentum along the ${\hat k}$ direction, and the three-vector $\vec{\kappa}$ is the preferred direction that breaks the Lorentz symmetry. This dispersion relation implies that the photon energy $\omega$ depends on the direction of the photon (${\hat k}$). A violation of Lorentz symmetry then generates a direction dependent vacuum refractive index, $n({\hat k}) \simeq 1 +  \vec{\kappa}\cdot{\hat k}$. In the MSME framework, anisotropies in the speed of light are bounded by three components of the coefficient $\tilde{\kappa}_{0+}$~\cite{mews011,mews012} which are then related to the components of $\vec{\kappa}$. In this work we used electron-photon scattering (Compton scattering) to study the energy dependence and the anisotropy of the vacuum refractive index. In a terrestrial laboratory the photon's three-momentum in a Compton scattering process changes direction due to the Earth's rotation, and thus any anisotropy in the vacuum dispersion would lead to sidereal variation in the kinematics of the scattering process. This work is similar to a measurement using Compton scattering at the European Synchrotron Radiation Facility (ESRF), but we use the Compton asymmetry instead of the kinematic edge (Compton edge) used in the ESRF study~\cite{graal1,graal2}. This work represents the first use of Compton scattering asymmetry to test Lorentz invariance.

Polarimeters based on Compton scattering are routinely used to continuously monitor the electron beam polarization at electron accelerators~\cite{SPEAR,LEP, Hera, Nikhef, LPOL, Bates, HallA1, HallA2, Escoffier, Friend, SLD1, SLD2, happex2,compton2}. These polarimeters involve measuring a known QED double-spin asymmetry in electron scattering from a photon beam of known polarization. The scattering asymmetry varies with the fraction of electron beam energy transferred to the scattered photon, with the largest
asymmetry occurring at the kinematic limit corresponding to the highest energy back-scattered photon. The Compton-scattered electrons and photons can be independently measured and analyzed to determine the polarization of the electron beam. Recently, the Compton polarimeter at Hall~C in  Jefferson Lab (JLab) demonstrated sub-1\% precision~\cite{compton2}. This high degree of precision of Compton polarimeters allows precise measurement of the vacuum refractive index of the Compton-scattered photon. By monitoring the sidereal variation of the vacuum refractive index we can also measure its isotropy. We report a
precise measurement of the vacuum refractive index for photons in the energy range of 9 - 46 MeV and MSME parameters
extracted from the sidereal variation of its deviation from unity.

\section{The Hall C Compton Polarimeter}
A parity-violating electron-scattering experiment in Hall~C at JLab, known as the \qweak experiment~\cite{qweakprl,qweaknim}, aims to test the Standard Model of particle physics by providing the first precision
measurement of the weak vector charge of the proton. In order to meet the high-precision requirement of the \qweak experiment, a new  Compton polarimeter was constructed in experimental Hall~C~\cite{qweaknim,compton1}.
A schematic of the Compton polarimeter in Hall~C at JLab is shown in Fig.~\ref{fig:fig1}.
The JLab Hall~C Compton polarimeter was designed to continuously monitor the beam polarization at high beam currents with better than 1\% statistical uncertainty, per hour. It consists of four 
identical dipole magnets forming a magnetic chicane that displaces a 1.16~GeV electron beam vertically downward by 57~cm. A high intensity ($\sim$ 1 - 3~kW) beam of $\sim$ 100\% circularly polarized photons is provided by an external low-gain \fp laser cavity which consists of a 85~cm long optical cavity with a gain between 100 and 300 coupled to a green (532~nm), continuous wave, 10 W laser (Coherent VERDI). The laser light is focused at the interaction region ($\sigma_{waist}\sim$~90~$\mu$m), and it is larger than the electron beam envelope ($\sigma_{x/y} \sim$~40~$\mu$m).    

After interacting with the photon target, the electron beam was deflected back to the nominal beamline with a second pair of dipole magnets. The third chicane magnet bends the primary beam by 10.1$^{\circ}$, and also separates the
Compton scattered electrons from the primary beam by up to 17~mm at the electron detector, placed just before the fourth dipole (Fig.~\ref{fig:fig1}). The electron detector consisted of a set of four planes of diamond micro-strip detectors. The close proximity of the detectors to the primary beam allowed them to capture $\sim$ 80\% of the spectrum of scattered electrons. The Compton electron spectrum is spread over 50 - 60 strips allowing a precise
measurement of its shape. The data analysis technique exploited track finding, the high granularity of the electron detector and its large acceptance to fit the shape of the measured asymmetry spectrum to the precisely calculable Compton asymmetry. The scattered photon had a maximum energy of approximately 46 MeV, which was detected in a calorimeter consisting of a 2$\times$2 matrix of PbWO$_4$ scintillating crystals attached to a single photo-multiplier tube. Details on the Compton polarimeter can be found in Ref.~\cite{qweaknim, compton1, compton2}. The results discussed here are from the electron detector only and were collected during the period between November, 2011 and May, 2012. 

The electron beam helicity was reversed at a rate of 960 Hz in a pseudo-random
sequence. The Compton laser was operated in 90 second cycles (60~s switched on and 30~s switched off).
The laser-off data were used to measure the background, which 
was subtracted from the laser-on yield for each electron helicity state.
The measured asymmetry was built from these yields as described in Ref.~\cite{compton1,compton2}. 
 A statistical precision of $<$~1\% per hour was routinely achieved. 

The electron beam polarization $P_e$ for a known photon polarization $P_{\gamma}$ was extracted by fitting the measured asymmetry to the $\mathcal{O}(\alpha)$ theoretical Compton asymmetry ($A_{\rm th}^j$) for fully polarized electrons and photon beams in the $j$-th strip, using
\begin{equation}
A_{\rm exp}^j = P_eP_{\gamma}A_{\rm th}^j.
\label{eq:pol}
\end{equation}
The electron polarization ($P_e$) and the non-integer strip number corresponding to the maximum displaced electrons ($j_{\rm max}$), i.e. the Compton edge, were the only free parameters in these fits. The theoretical Compton asymmetry $A_{\rm th}(\rho)$ in Eq.~\ref{eq:pol} was calculated as a function of the dimensionless variable
\begin{equation}
\rho=\frac{E_{\gamma}}{E_{\gamma}^{\rm max}} \approx \frac{E_e^{\rm beam}-E_e}{E_e^{\rm beam}-E_e^{\rm min}},
\label{eq:rho}
\end{equation}
where $E_{\gamma}$ is the energy of a back-scattered photon,  $E_{\gamma}^{\rm max}$ is the maximum allowed 
photon energy, and $E_e$, $E_e^{\rm min}$, and $E_e^{\rm beam}$ are the scattered electron energy, its minimum value and the electron beam energy, respectively. $A_{\rm th}^n$ was related to $A_{\rm th}(\rho)$  by mapping $\rho$ to the strip number using knowledge of the magnetic field in the third dipole, the geometry of the chicane, the strip pitch and an initial estimate of $j_{\rm max}$ from the scattered electron spectrum.

\section{Refractive Index of Free Space From Compton Scattering}
The variation in speed of light can be represented as a variation in the refractive index of vacuum. If the speed of light was constant with respect to a parameter such as photon energy ($\omega$), the refractive index of vacuum ($n(\omega)$) is a constant equal to unity. The search for a deviation of the refractive index of vacuum from unity as a function photon energy or photon direction is then a search for violation of Lorentz symmetry. For Compton scattering of ultra-relativistic electrons in vacuum, the effect of any deviation of the refractive index from unity is evaluated by re-writing the energy-momentum conservation for $n\approx 1$ (to $\mathcal{O}[(n-1)^2]$)~\cite{vahprl};
\begin{equation}
E_{0}x - \omega(1+x+\gamma^2 \theta^2) + 2\omega_0(1 - \frac{\omega}{E_0})\gamma^2(n-1) =0, 
\label{consveq}
\end{equation}
where the electrons have initial energy $E_0$ (1.16~GeV) and mass $m_e$ and the photons have initial and 
final energy $\omega_0$ ($2.32$ eV), $\omega$ ($9 - 46$ MeV) and initial and final angle $\theta_0, \theta$ respectively, and $x = 4\gamma \omega_0 \sin^{2}{(\frac{\theta_0^2}{2})}/m_e$. 
 As seen in Eq.~\ref{consveq} any deviation of the refractive index 
from unity is amplified by the square of the Lorentz boost. This makes Compton scattering very 
sensitive to tiny deviations of the refractive index from unity~\cite{vahplb}, and the sensitivity grows 
rapidly with increasing electron beam energy.

In the JLab Hall~C Compton polarimeter, the scattered electrons are momentum analyzed by dipole-3 (see Fig.~\ref{fig:fig1}) and detected in the position sensitive diamond micro-strip detector, which measures the deflection of the scattered electrons with respect to the unscattered electrons. The effect of any deviation of the refractive index from unity can be incorporated into the Compton cross section and longitudinal asymmetry by modifying the dimensionless variable $\rho$ as~\cite{vahplb}, 
\begin{equation}
\rho(n) = \rho \left[\frac{1 + \frac{2\gamma^2(n-1)(1+\gamma^2\theta^2)}{(1+x+\gamma^2\theta^2)^2}}{1 + \frac{2\gamma^2(n-1)}{(1+x)^2}}\right].
\label{eq:rhon}
\end{equation}

The above equation can be written up to $\mathcal{O}[(n-1)^2]$ as;
\begin{equation}
\rho(n) = \rho \left[1 + 2\gamma^2(n-1)f(x,\theta)\right],
\label{eq:rhon2}
\end{equation}
where the kinematic function $f(x,\theta)$ is given by,
\begin{equation}
f(x,\theta) = \frac{(1+\gamma^2\theta^2)(1+x)^2 - (1+x^2+\gamma^2 \theta^2)^2}{(1+x+\gamma^2\theta^2)^2(1+x)^2}
\end{equation}

Moreover, the direction dependent vacuum refractive index is written as $n({\hat k}) - 1 \simeq \vec{\kappa}\cdot{\hat k}$, and the variable $\rho(n)$ is also direction dependent as:
\begin{equation}
\rho(n) \simeq \rho \left[1 + 2\gamma^2f(x,\theta)\vec{\kappa}\cdot{\hat k}\right],
\label{eq:rhon3}
\end{equation}

\begin{figure}[hbtp!]
\includegraphics*[width=9.0cm]{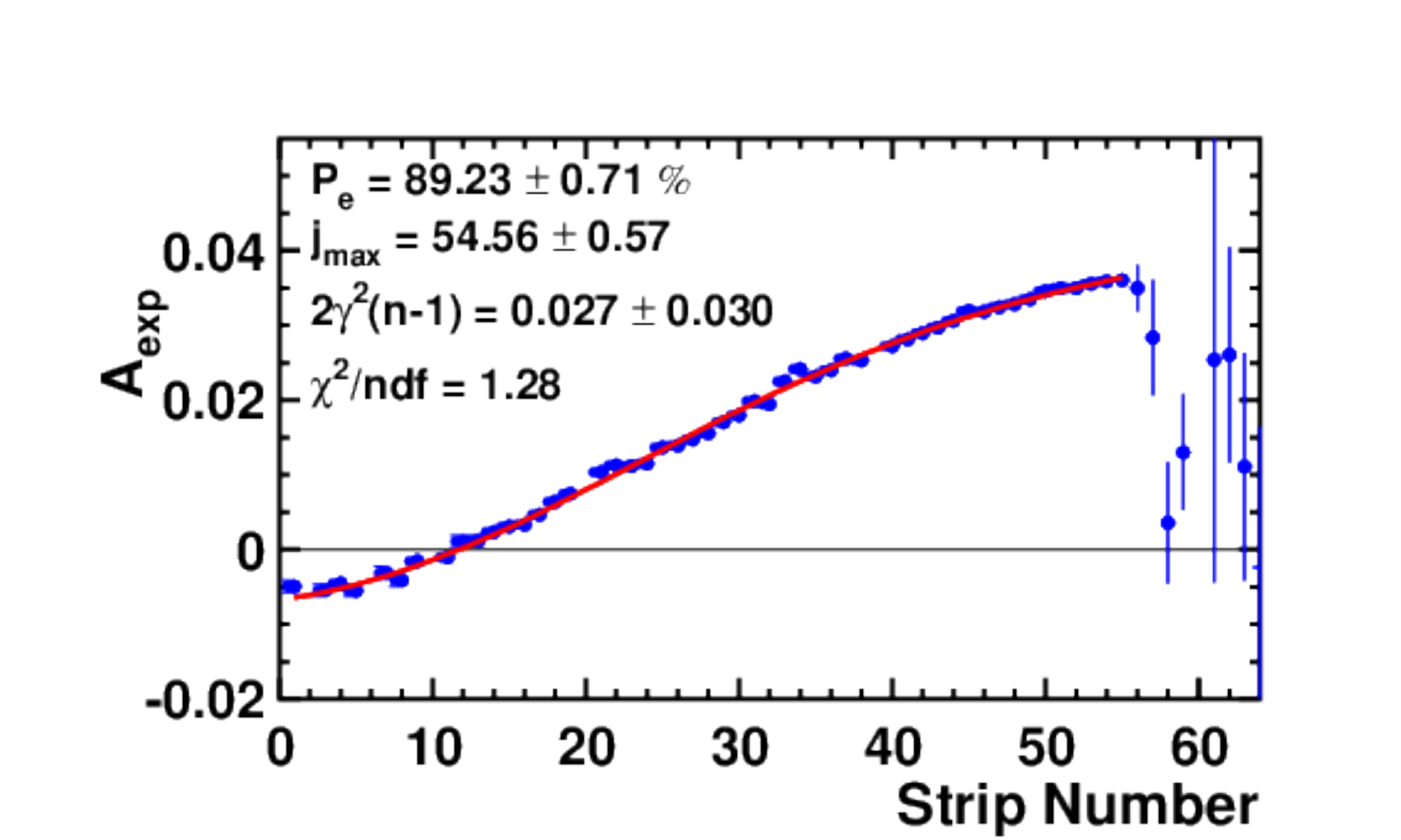}
\caption[]{The measured Compton asymmetry (background-subtracted) from a single detector plane plotted versus detector strip number for a typical hour-long run. Statistical uncertainties only are shown. The solid red line is a 3-parameter fit to the  $\mathcal{O}[(n-1)^2]$ calculated asymmetry.} 
\label{fig:fig2}
\end{figure}

The theoretical asymmetry calculated as a function of the modified $\rho(n)$ (Eq.~\ref{eq:rhon}) and mapped to 
each strip was fit to the measured asymmetry with three free parameters; the electron beam polarization, $P_e$, the strip location of the Compton edge, $j_{max}$ and the quantity $2\gamma^2(n-1)$. As mentioned earlier the parameters $P_e$ and $j_{max}$ were used in the regular analysis for extracting the electron beam polarization as described in Ref.~\cite{compton2,compton1}. The additional parameter $2\gamma^2(n-1)$ along with the modified dimensionless variable $\rho(n)$ is used, only in the analysis described here, to extract the deviation of the refractive index from unity. The parameters $P_e$ and $j_{max}$ obtained from the 3-parameter fit were within one standard deviation of the values obtained in the 2-parameter fit. The $\chi^2$ per degree-of-freedom of the 3-parameter fit, considering statistical uncertainties only, ranges between 0.8 and 1.6 for 50~--~60 degrees of freedom. A typical measured Compton asymmetry for an hour long run is shown in Fig.~\ref{fig:fig2}. The 3-parameter fit of the measured asymmetry to the theoretical asymmetry is shown by the solid curve. The systematic uncertainties of the Compton polarimeter were determined to be 0.6\%~\cite{compton2} when using the 2-parameter fit of the asymmetry. The 3-parameter fit used in this analysis led to an additional contribution to the systematic uncertainties. The additional uncertainty was estimated from the variation in the parameters $P_e$ and $j_{max}$ between the 2 and 3 parameter fits. The quadrature sum of the instrumental and fit uncertainties resulted in a net systematic uncertainty of 0.96\%.     

\begin{figure}[hbtp!]
\includegraphics*[width=9.0cm]{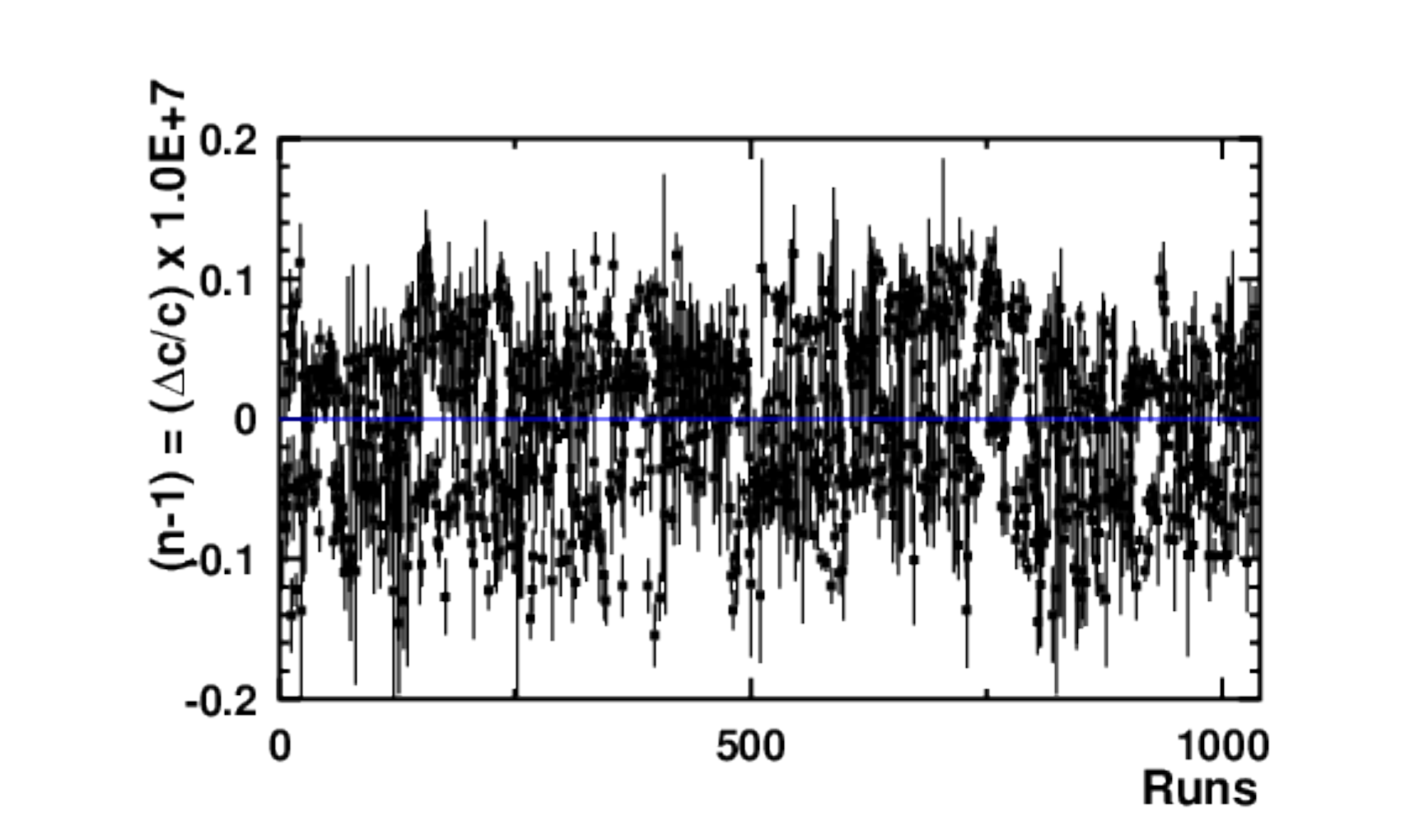} 
\caption[]{The deviation from unity ($n-1$) of the vacuum refractive index, extracted from the fit of the measured asymmetry to the calculated asymmetry, as a function of run number. The total uncertainty given by the quadrature sum of the statistical and systematic uncertainties are shown.}
\label{fig:fig3}
\end{figure}

\begin{figure}[hbtp!]
\includegraphics*[width=8.5cm]{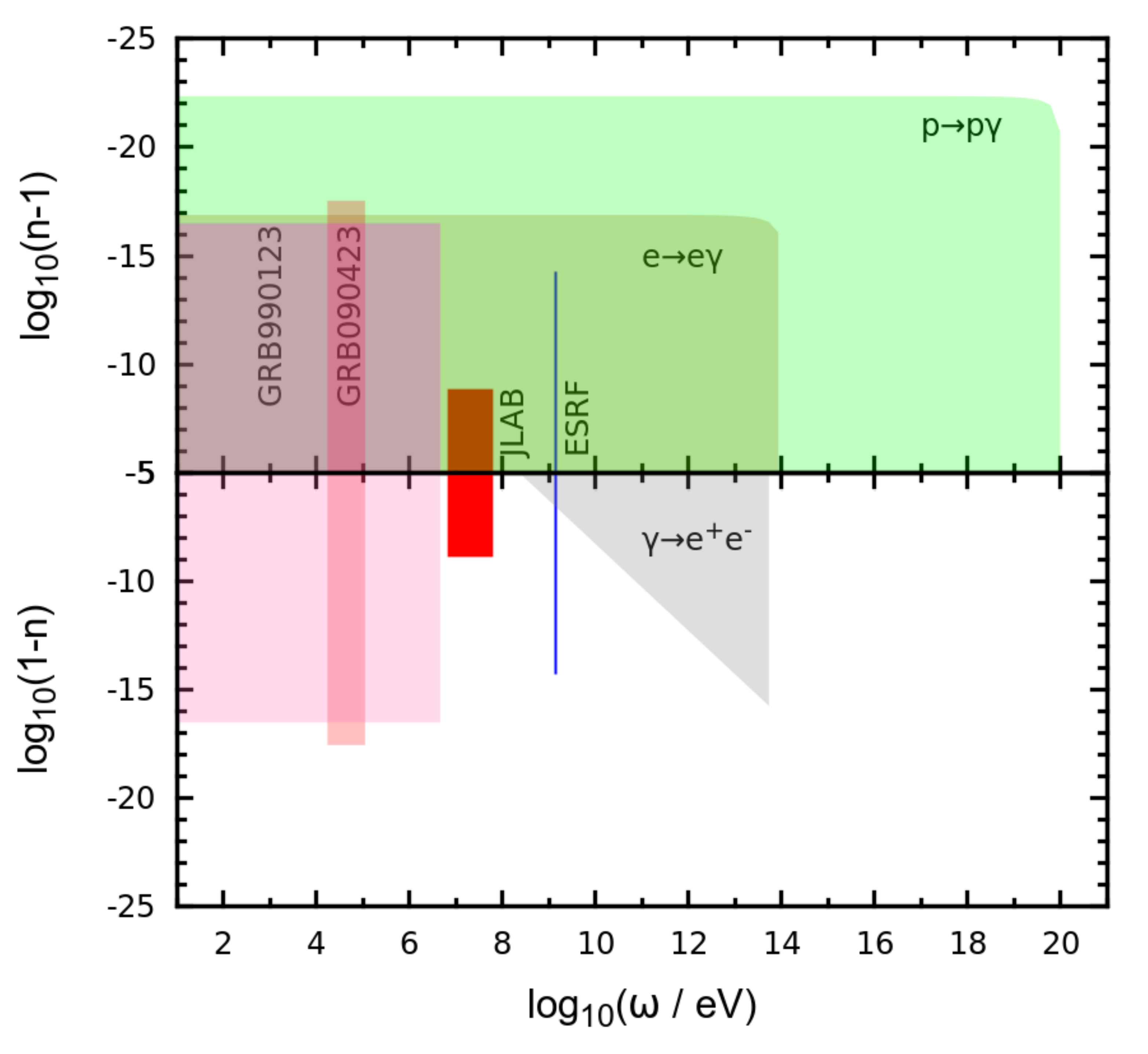} 
\caption[]{The constraints on the energy dependence of the vacuum refractive index imposed by existing measurements. Shaded areas are excluded regions. The results from the JLab Hall~C Compton polarimeter are shaded red.}
\label{fig:fig4}
\end{figure}

The value of $n-1$ (deviation of refractive index from unity) for the entire six month period of the \qweak run-2, for each 1-hr long run is shown in Fig.~\ref{fig:fig3}. The distribution of $n-1$ for all of run-2 was fit to a Gaussian distribution. The $\chi^2$ per degree-of-freedom of the fit was 1.44. The width parameter of the Gaussian distribution $(6.8 \times 10^{-9})$ was consistent with the mean $(0.7 \times 10^{-8})$ of the the distribution of total uncertainty  of the entire data set. From the mean and the width of the Gaussian fit to the distribution of $n-1$ we obtain $n = 1 + (0.1 \pm 6.8) \times 10^{-9}$. This excludes deviation of the vacuum refractive index from unity, larger than $1.4\times 10^{-8}$ at the 95\% confidence level, for $9 - 46$ MeV photons. It is the most stringent bound, to date, on the refractive index being less than one, for photon energies of $9 - 46$ MeV. Our results along with all existing bounds on the energy dependence of the refractive index of vacuum are shown in Fig~\ref{fig:fig4}. For example, in a dispersive medium whose refractive index is less than one, photons would quickly decay via pair production, while in a medium whose refractive index is greater than one, charged particles would rapidly decay by vacuum \v{C}erenkov radiation. Therefore, the highest energy photons and charged particles detected in cosmic rays provide the most stringent constrains on the deviation of refractive index of vacuum from unity~\cite{coleman97}. The highest energy proton detected ($10^{20}$eV \cite{wdowczyk89}) excludes the region shaded as light green (Fig.~\ref{fig:fig4}) and marked by $p\rightarrow p\gamma$. Similarly, the highest energy electron detected by the HESS collaboration is about $100$~TeV~\cite{hess} and it leads to the exclusion region marked as $e\rightarrow e\gamma$. The highest energy cosmic photons detected is $22$TeV~\cite{hegra} and it leads to the exclusion region shown in gray and marked by $\gamma \rightarrow e\bar{e}$.  The exclusion regions named GRB990123 \cite{grb99} and GRB090423 \cite{grb09}  are obtained from  gamma ray bursts using their distance and the time delay ($\Delta t$) between the hard and soft parts of the spectra. This exclusion follows from the constraint that $|n - 1| < c\Delta t/D$, where D is the distance to the gamma ray burst. Also shown is the exclusion from the another Compton scattering measurement at the ESRF using $6$~GeV electrons~\cite{graal1,graal2}.

\subsection{Sidereal Variation and Constraints on SME Parameters}
Any hypothetical violation of Lorentz symmetry generates a 
direction dependent vacuum refractive index. By searching for a 
sidereal modulation in the vacuum refractive index extracted from the 
Compton polarimetry data shown in Fig~\ref{fig:fig3}, we can constrain the leading MSME coefficient responsible for direction dependence of the speed of light, $\tilde{\kappa}_{0+}$~\cite{mews011,mews012}. The directional dependence of the vacuum refractive index can be written as $n({\hat k}) -1 \simeq \vec{\kappa}\cdot{\hat k}$, where
${\hat k}$ is the direction of the back scattered photon in the Compton scattering process and $\vec{\kappa}$ is the 3-vector along the hypothetical preferred direction that breaks the Lorentz symmetry. The three components of MSME coefficient $\tilde{\kappa}_{0+}$ have been incorporated into $\vec{\kappa}$ as: $\vec{\kappa} \equiv \left< \left( \tilde{\kappa}_{0^+}^{23} \right), \left( \tilde{\kappa}_{0^+}^{31} \right), \left( \tilde{\kappa}_{0^+}^{12} \right) \right> \equiv \left<\kappa_X, \kappa_Y, \kappa_Z \right>$. 

\begin{figure}[hbtp!]
\includegraphics*[width=9.0cm]{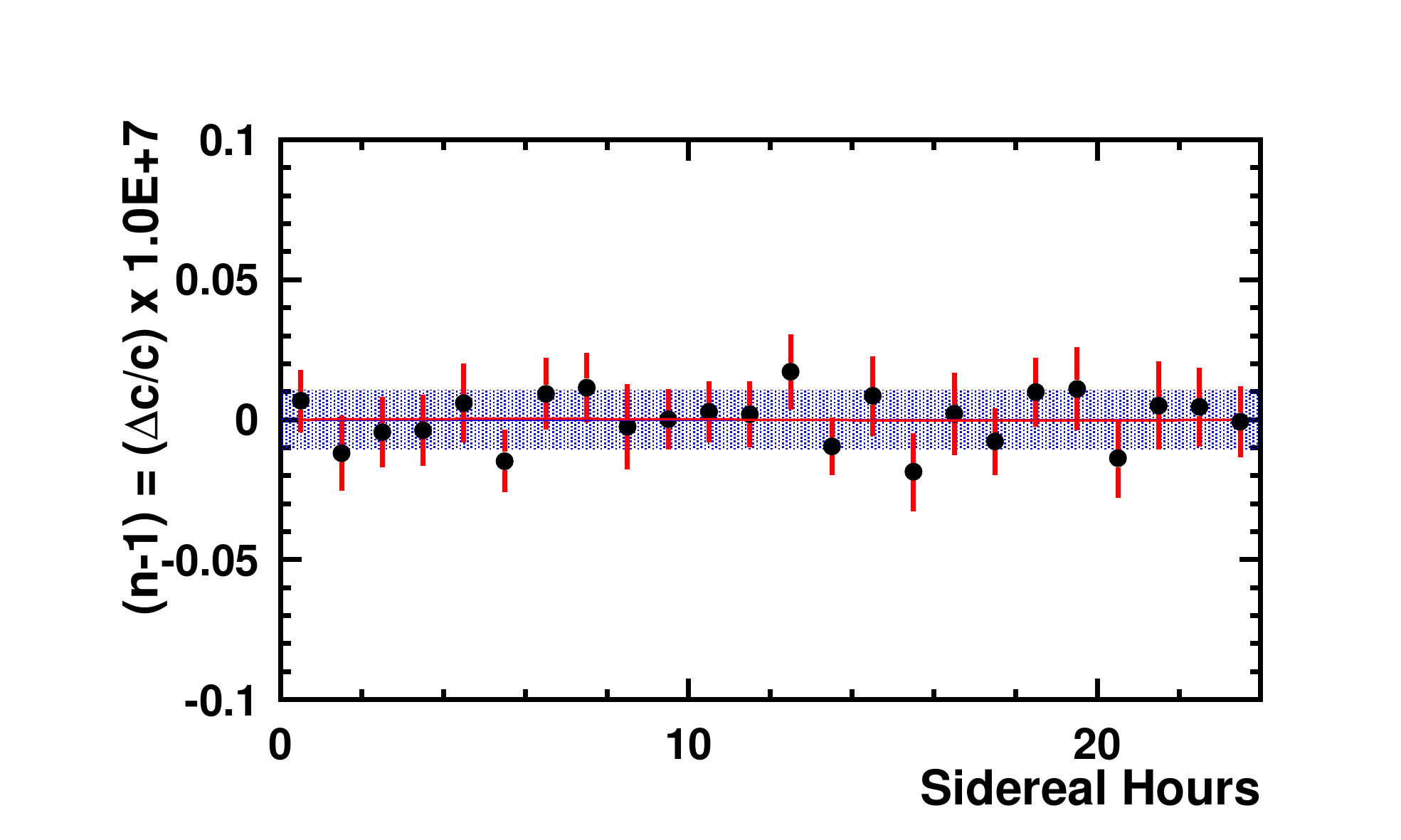} 
\caption[]{The deviation of the vacuum refractive index from unity as a function of sidereal time, with the entire data set modulo one sidereal day (24 bins). The quadrature sum of statistical and systematic uncertainties is shown.}
\label{fig5}
\end{figure}

The MSME coefficients are typically specified in the Sun-centered inertial frame~\cite{litests}, where the $\hat{z}$ direction is parallel to the Earth's axis of rotation, with the Earth rotating at a frequency $\Omega \approx 2\pi$/(23~hr~56~min) about $\hat{z}$. The direction of the scattered 
photon beam at JLab Hall~C is given by $\hat{k} = \left<0.13 cos(\Omega t), 0.87 sin(\Omega t), 0.48 \right>$. Hence the vacuum refractive index can be written as:
\begin{equation}
n - 1 \simeq  \vec{\kappa}\cdot{\hat k} \simeq \left( 0.87 \sqrt{\kappa_X^2 + \kappa_Y^2} sin(\Omega t)\right)
\label{eq:np}
\end{equation}
Here an irrelevant phase has been neglected. To search for a sidereal modulation, the entire data set was folded modulo one sidereal day and divided into 1-hr bins (24 bins) as shown in Fig.~\ref{fig5}. The data were then fit to a sinusoidal function with a frequency of $\Omega$. The amplitude of the hypothetical sidereal oscillation was found to be $(0.3 \pm 3.8) \times 10^{-10}$. This  gives a 95\% upper bound on the sidereal oscillation of $7.5 \times 10^{-10}$, yielding a limit of $\sqrt{\kappa_X^2 + \kappa_Y^2} <  8.6 \times 10^{-10}$. In the MSME notation including the electron coefficients for generality, this gives at 95\% confidence level:     
\begin{equation}
\sqrt{[2c_{TX} - (\tilde{\kappa}_{0+})^{YZ}]^2 + [2c_{TY} - (\tilde{\kappa}_{0+})^{ZX}]^2} < 8.6 \times 10^{-10},
\end{equation}
This bound is several order of magnitudes worse than the current best limit~\cite{graal1,graal2}. However, it demonstrates the feasibility of a new technique which will yield much more stringent limits in the future when the Compton polarimeters at JLab are operated at a higher electron beam energy of 11 GeV.

\section{Conclusion}
We have used the Compton scattering asymmetry of a 1.16~GeV electron beam at JLab to measure the vacuum refractive index at a photon energy range of 9 - 46 MeV. We have also used the sidereal variation of these measurements to extract the MSME parameters for directional dependence of vacuum refractive index. These results demonstrate that it is feasible to use Compton scattering asymmetry to search for Lorentz symmetry violation and the sensitivity of such measurements increase with the electron beam energy. Our results improve the bounds on the energy dependence of the vacuum refractive index in the 9 - 46 MeV photon energy range. Given the low electron beam energy, the bounds on the direction dependence of the vacuum refractive index from our results are not competitive. However, future parity violating electron scattering experiments at JLab will use a 11~GeV beam with even higher precision Compton polarimeters~\cite{future1,future2}. The same technique applied to these experiments should provide more stringent constraints on Lorentz symmetry. 

\section*{Acknowledgments}
This work was funded by the U.S. Department of Energy, including contract 
\#DE-FG02-07ER41528 and \#AC05-06OR23177 under which Jefferson Science Associates, LLC operates Thomas Jefferson National Accelerator Facility, and by the U.S. National Science Foundation and  the Natural Sciences and Engineering Research Council of Canada (NSERC). We wish to thank the staff of JLab, TRIUMF, and MIT-Bates, for their vital support. One of the authors (P. M) would like to thank W. Deconinck, V. Gharibyan and F. Guo for having detailed conversations regarding the experimental apparatus and theory side of this paper.

\end{document}